# A note on proper affine symmetry in Kantowski-Sachs and Bianchi type III space-times

Ghulam Shabbir and Nisar Ahmed

## Abstract

We investigate proper affine symmetry for the Kantowski-Sachs and Bianchi type III space-times by using holonomy and decomposability, the rank of the 6×6 Riemann matrix and direct integration techniques. It is shown that the very special classes of the above space-times admit proper affine vector fields.

## 1. INTRODUCTION

The aim of this paper is to find the existence of proper affine vector fields in Kantowski-Sachs and Bianchi type III space-times by using holonomy and decomposability, the rank of the 6×6 Riemann matrix and direct integration techinques. The affine vector field which preserves the geodesic structure and affine parameter of a space-time carries significant information and interest in the Einstein's theory of general relativity. It is therefor important to study this symmetry. Let $(M, g)$ be a space-time with $M$ a smooth connected Hausdorff four dimensional manifold and $g$ a smooth metric of Lorentz signature (-, +, +, +) on $M$. The curvature tensor associated with g, through Levi-Civita connection, is denoted in component form by $R^a{}_{bcd}$. The usual covariant, partial and Lie derivatives are denoted by a semicolon, a comma and the symbol $L$, respectively. Round and square brackets denote the usual symmetrization and skew-symmetrization, respectively. The space-time $M$ will be assumed nonflat in the sense that the Riemann tensor does not vanish over any non-empty open subset of $M$.

A vector field $X$ on $M$ is called an affine vector field if it satisfies

$$X_{a;bc} = R_{abcd} X^d \tag{1}$$

or equivalently,

$$X_{a,bc} - \Gamma^f_{ac} X_{f,b} - \Gamma^f_{bc} X_{a,f} - \Gamma^e_{ab} X_{e,c} + \Gamma^e_{ab}\Gamma^f_{ec} X_f - (\Gamma^e_{ab})_{,c} X_e - \Gamma^f_{ab}\Gamma^e_{cf} X_e$$
$$+ \Gamma^e_{fb}\Gamma^f_{ca} X_e + \Gamma^e_{af}\Gamma^f_{bc} X_e = R_{abcd} X^d.$$

If one decomposes $X_{a;b}$ on $M$ into its symmetric and skew-symmetric parts

$$X_{a;b} = \frac{1}{2} h_{ab} + F_{ab} \qquad (h_{ab} = h_{ba},\ F_{ab} = -F_{ba}) \tag{2}$$

then equation (1) is equivalent to

$$(i)\ h_{ab;c} = 0 \quad (ii)\ F_{ab;c} = R_{abcd} X^d \quad (iii)\ F_{ab;c} X^c = 0. \tag{3}$$

If $h_{ab} = 2cg_{ab},\ c \in R,$ then the vector field $X$ is called homothetic (and *Killing* if $c = 0$). The vector field $X$ is said to be proper affine if it is not homothetic vector field and also $X$ is said to be proper homothetic vector field if it is not Killing vector field on $M$ [2]. Define the subspace $S_p$ of the tangent space $T_pM$ to $M$ at $p$ as those $k \in T_pM$ satisfying

$$R_{abcd} k^d = 0. \tag{4}$$

## 2. Affine Vector Fields

Suppose that $M$ is a simple connected space-time. Then the holonomy group of $M$ is a connected Lie subgroup of the idenity component of the Lorentz group and is thus characterized by its subalgebra in the Lorentz algebra. These have been labeled into fifteen types $R_1 - R_{15}$ [1]. It follows from [2] that the only such space-times which could admit proper affine vector fields are those which admit nowhere zero covariantly constant second order symmetric tensor field $h_{ab}$. This forces the holonomy type to be either $R_2$, $R_3$, $R_4$, $R_6$, $R_7$, $R_8$, $R_{10}$, $R_{11}$ or $R_{13}$ [6]. A study of the affine vector fields for the above holonomy types can be found in [2]. It follows from [3] that the rank of the $6 \times 6$ Riemann matrix of the above space-times which have holonomy type $R_2$, $R_3$, $R_4$, $R_6$, $R_7$, $R_8$, $R_{10}$, $R_{11}$ or $R_{13}$ is atmost three. Hence for studying affine vector fields we are interested in those cases when the rank of the $6\times 6$ Riemann matrix is less than or equal to three.

## 3. MAIN RESULTS

Consider the space-times in the usual coordinate system $(t,r,\mathbf{q},\mathbf{f})$ with line element [4]

$$ds^2 = -dt^2 + A(t)dr^2 + B(t)[d\mathbf{q}^2 + f^2(\mathbf{q})d\mathbf{f}^2], \qquad (5)$$

where $A$ and $B$ are no where zero functions of $t$ only. For $f(\mathbf{q}) = \sin \mathbf{q}$ or $f(\mathbf{q}) = \sinh \mathbf{q}$ the above space-time (5) become Kantowski-Sachs or Bianchi type III space-times, respectively. The above space-time admits four independent Killing fields which are [5]

$$\frac{\partial}{\partial r}, \quad \frac{\partial}{\partial \mathbf{f}}, \quad \cos\mathbf{f}\frac{\partial}{\partial \mathbf{q}} - \frac{f'}{f}\sin\mathbf{f}\frac{\partial}{\partial \mathbf{f}}, \quad \sin\mathbf{f}\frac{\partial}{\partial \mathbf{q}} + \frac{f'}{f}\cos\mathbf{f}\frac{\partial}{\partial \mathbf{f}}, \qquad (6)$$

where prime denotes the derivative with respect to $\mathbf{q}$. The non-zero independent components of the Riemann tensor are

$$R_{0101} = \frac{\dot{A}^2 - 2\ddot{A}A}{4A} \equiv \mathbf{a}_1, \qquad R_{0202} = \frac{\dot{B}^2 - 2\ddot{B}B}{4B} \equiv \mathbf{a}_2,$$

$$R_{0303} = f^2(\mathbf{q})\left(\frac{\dot{B}^2 - 2\ddot{B}B}{4B}\right) \equiv \mathbf{a}_3, \qquad R_{1212} = \frac{\dot{A}\dot{B}}{4} \equiv \mathbf{a}_4, \qquad (7)$$

$$R_{1313} = f^2(\mathbf{q})\left(\frac{\dot{A}\dot{B}}{4}\right) \equiv \mathbf{a}_5, \qquad R_{2323} = f^2(\mathbf{q})\left(\frac{\dot{B}^2 + 4B}{4}\right) \equiv \mathbf{a}_6,$$

where dot denotes the derivative with respect to $t$. One can write the curvature tensor with components $R_{abcd}$ at $p$ as a $6\times 6$ symmetric matrix

$$R_{abcd} = diag(\mathbf{a}_1, \mathbf{a}_2, \mathbf{a}_3, \mathbf{a}_4, \mathbf{a}_5, \mathbf{a}_6), \qquad (8)$$

where $\mathbf{a}_1, \mathbf{a}_2, \mathbf{a}_3, \mathbf{a}_4, \mathbf{a}_5$ and $\mathbf{a}_6$ are real functions of $t$ on $M$. As mentioned in section 2, the space-times which can admit proper affine vector fields have holonomy type $R_2$, $R_3$, $R_4$, $R_6$, $R_7$, $R_8$, $R_{10}$, $R_{11}$ or $R_{13}$ and the rank of the $6\times 6$ Riemann matrix is atmost three. Hence, we are only interested in those cases when the rank of the $6\times 6$ Riemann matrix is less than or equal to three (excluding the flat cases). Thus there exist the following possibilities:

(A1)　　Rank $= 3$, $\dot{A}\neq 0$, $\dot{B}\neq 0$, $\dot{A}^2 - 2\ddot{A}A \neq 0$, $\dot{B}^2 - 2\ddot{B}B = 0$, $\dot{B}^2 + 4B = 0$.

(A2)　　Rank $= 3$, $\dot{A}= 0$, $\dot{B}\neq 0$, $\dot{B}^2 - 2\ddot{B}B \neq 0$, $\dot{B}^2 + 4B \neq 0$.

(A3)　　Rank $= 3$, $\dot{A}\neq 0$, $\dot{B}\neq 0$, $\dot{A}^2 - 2\ddot{A}A = 0$, $\dot{B}^2 - 2\ddot{B}B = 0$, $\dot{B}^2 + 4B \neq 0$.

(A4)　　Rank $= 2$, $\dot{A}\neq 0$, $\dot{B}= 0$, $\dot{A}^2 - 2\ddot{A}A \neq 0$.

(A5)　　Rank $= 2$, $\dot{A}\neq 0$, $\dot{B}\neq 0$, $\dot{A}^2 - 2\ddot{A}A = 0$, $\dot{B}^2 - 2\ddot{B}B = 0$, $\dot{B}^2 + 4B = 0$.

(A6)　　Rank $= 1$, $\dot{A}= 0$, $\dot{B}= 0$.

(A7)　　Rank $= 1$, $\dot{A}= 0$, $\dot{B}\neq 0$, $\dot{B}^2 - 2\ddot{B}B = 0$, $\dot{B}^2 + 4B \neq 0$.

(A8)　　Rank $= 1$, $\dot{A}\neq 0$, $\dot{B}= 0$, $\dot{A}^2 - 2\ddot{A}A = 0$.

We consider each case in turn.

**Case A1**

In this case $\dot{A}\neq 0$, $\dot{B}\neq 0$, $\dot{A}^2 - 2\ddot{A}A \neq 0$, $\dot{B}^2 - 2\ddot{B}B = 0$, $\dot{B}^2 + 4B = 0$, the rank of the $6\times 6$ Riemann matrix is three and there exist no non-trivial solutions of equation (4). Equations $\dot{B}^2 - 2\ddot{B}B = 0$ and $\dot{B}^2 + 4B = 0 \Rightarrow$ $B(t) = -t^2 + c_1 t - \frac{1}{4}c_1^2$, where $c_1 \in R$. The line element can be written in the form

$$ds^2 = -dt^2 + A(t)dr^2 + \left(-t^2 + c_1 t - \frac{1}{4}c_1^2\right)\left[d\boldsymbol{q}^2 + f^2(\boldsymbol{q})d\boldsymbol{f}^2\right] \quad (9)$$

Substituting the above information in (1) and after some calculation one finds that in this case affine vector fields are Killing which are given in equation (6).

**Case A2**

In this case $\dot{A}= 0$, $\dot{B}\neq 0$, $\dot{B}^2 - 2\ddot{B}B \neq 0$, $\dot{B}^2 + 4B \neq 0$ and the rank of the $6\times 6$ Riemann matrix is three. Equation $\dot{A}= 0 \Rightarrow A = c_1$, where $c_1 \in R\setminus\{0\}$.

Here, there exists a covariantly constant vector field $r_a$, which is a unique solution (up to a multiple) of equation (4) i.e. $r_{a;b} = 0$ and consequently the Ricci identity implies that $R_{abcd}r^d = 0$. The line element can, after a suitable rescaling of $r$ be written in the form

$$ds^2 = dr^2 + [-dt^2 + B(t)(d\mathbf{q}^2 + f^2(\mathbf{q})d\mathbf{f}^2)], \qquad (10)$$

The space-time is clearly 1+3 decomposable. Affine vector fields in this case [2] are of the form

$$X = (rc_1 + c_2)\frac{\partial}{\partial r} + X', \qquad (11)$$

where $c_1, c_2 \in R$ and $X'$ is a homothetic vector field in the induced geometry on each of the three dimensional submanifolds of constant $r$. The completion of case A2 requires to find the homothetic vector fields in the induced geometry on the submanifolds of constant $r$. The induced metric $g_{ab}$ (where $a,b = 0,2,3$) has non zero components given by

$$g_{00} = -1, \qquad g_{22} = B(t), \qquad g_{33} = B(t)f^2(\mathbf{q}). \qquad (12)$$

A vector field $X'$ is called a homothetic vector field if it satisfies

$$\mathcal{L}_{X'} g_{ab} = 2c g_{ab}, \qquad c \in R. \qquad (13)$$

One can expand (13) and using (12) to get

$$X^0{}_{,0} = c, \qquad (14)$$

$$X^0{}_{,2} - B(t)X^2{}_{,0} = 0, \qquad (15)$$

$$X^0{}_{,3} - B(t)f^2(\mathbf{q})X^3{}_{,0} = 0, \qquad (16)$$

$$\dot{B}(t)X^0 + 2B(t)X^2{}_{,2} = 2cB(t), \qquad (17)$$

$$X^2{}_{,3} + f^2(\mathbf{q})X^3{}_{,2} = 0, \qquad (18)$$

$$\dot{B}(t)f(\mathbf{q})X^0 + 2B(t)f'(\mathbf{q})X^2 + 2B(t)f(\mathbf{q})X^3{}_{,3} = 2cB(t)f(\mathbf{q}). \qquad (19)$$

Equations (14), (15) and (16) to give

$$X^0 = ct + C^0(\mathbf{q},\mathbf{f}), \quad X^2 = C_q^0(\mathbf{q},\mathbf{f})\int \frac{dt}{B(t)} + C^1(\mathbf{q},\mathbf{f}),$$

$$X^3 = \frac{1}{f^2(\mathbf{q})} C_f^0(\mathbf{q},\mathbf{f})\int \frac{dt}{B(t)} + C^2(\mathbf{q},\mathbf{f}), \tag{20}$$

where $C^0(\mathbf{q},\mathbf{f})$, $C^1(\mathbf{q},\mathbf{f})$, and $C^2(\mathbf{q},\mathbf{f})$ are functions of integration. If one proceeds further, after a straightforward calculation one can find that the proper homothetic vector field exists if and only if $B(t) = (c_2 t + c_3)^2$, where $c_2, c_3 \in R(c_2 \neq 0)$. Substituting the above information into (7), one finds that the rank of the 6×6 Riemann matrix is reduced to two, thus giving a contradiction. So the only homothetic vector fields in the induced geometry are the Killing vector fields which are given by

$$X^0 = 0, \quad X^2 = c_4 \cos \mathbf{f} + c_5 \sin \mathbf{f},$$

$$X^3 = -c_4 \frac{f'}{f} \sin \mathbf{f} + c_5 \frac{f'}{f} \cos \mathbf{f} + c_6, \tag{21}$$

where $c_4, c_5, c_6 \in R$. Affine vector fields in this case are given by use of (11) and (21) as

$$X^0 = 0, \quad X^1 = (rc_1 + c_2), \quad X^2 = c_4 \cos \mathbf{f} + c_5 \sin \mathbf{f},$$

$$X^3 = -c_4 \frac{f'}{f} \sin \mathbf{f} + c_5 \frac{f'}{f} \cos \mathbf{f} + c_6. \tag{22}$$

One can write the above equation (22) after subtracting Killing vector fields as

$$X = (0, r, 0, 0). \tag{23}$$

Clearly, the above space-time admits proper affine vector field.

**Case A3**

In this case we have $\dot{A} \neq 0$, $\dot{B} \neq 0$, $\dot{A}^2 - 2\ddot{A} A = 0$, $\dot{B}^2 - 2\ddot{B} B = 0$ and $\dot{B}^2 + 4B \neq 0$. Equations $\dot{A}^2 - 2\ddot{A} A = 0$ and $\dot{B}^2 - 2\ddot{B} B = 0$ implies $A(t) = (at + b)^2$ and $B(t) = (ct + d)^2$, where $a, b, c, d \in R(a, c \neq 0)$. We first suppose that $a \neq c$ and $b \neq d$. The rank of the 6×6 Riemann matrix is three, and there exists a unique solution (up to a multiple) $t_a = t_{,a}$ of equation (4) but $t_a$ is not covariantly constant. The line element is

$$ds^2 = -dt^2 + (at+b)^2 dr^2 + (ct+d)^2 (d\mathbf{q}^2 + f^2(\mathbf{q})d\mathbf{f}^2). \tag{24}$$

Substituting the above information into affine equations (1) and after some calculation one finds that affine vector fields in this case are

$$X^0 = 0, \quad X^1 = c_7, \quad X^2 = c_4 \cos \mathbf{f} + c_5 \sin \mathbf{f},$$
$$X^3 = -c_4 \frac{f'}{f} \sin \mathbf{f} + c_5 \frac{f'}{f} \cos \mathbf{f} + c_6, \tag{25}$$

provided that $ad - bc \neq 0$ and $c_4, c_5, c_6, c_7 \in R$. Affine vector fields in this case are Killing vector fields.

Now consider the case if $ad - bc = 0$ than affine vector fields in this case are

$$X^0 = c_2 t + c_3, \quad X^1 = c_7, \quad X^2 = c_4 \cos \mathbf{f} + c_5 \sin \mathbf{f},$$
$$X^3 = -c_4 \frac{f'}{f} \sin \mathbf{f} + c_5 \frac{f'}{f} \cos \mathbf{f} + c_6, \tag{26}$$

where $c_2, c_3, c_4, c_5, c_6, c_7 \in R$. One can write the above equation (26) after subtracting Killing vector fields as

$$X = (c_2 t + c_3, 0, 0, 0). \tag{27}$$

Clearly, the above space-time admits proper affine vector fields.

Now consider the case $a = c$ and $b = d$. The line element takes the form

$$ds^2 = -dt^2 + (at+b)^2 (dr^2 + d\mathbf{q}^2 + f^2(\mathbf{q})d\mathbf{f}^2). \tag{28}$$

The above space-time (28) admits proper affine vector fields which are given in equation (27).

**Case A4**

In this case we have $\dot{A} \neq 0$, $\dot{A}^2 - 2\ddot{A}A \neq 0$, $\dot{B} = 0$ and the rank of the 6×6 Riemann matrix is two. Equation $\dot{B} = 0$ implies that $B(t) = c_1$, where $c_1 \in R \setminus \{0\}$. There exists no non trivial solution of equation (4). The line element takes the form

$$ds^2 = -dt^2 + A(t)dr^2 + c_1 (d\mathbf{q}^2 + f^2(\mathbf{q})d\mathbf{f}^2). \tag{29}$$

The above space-time is clearly 2+2 decomposable and affine vector fields in this case [2] take the form

$$X = X_1 + X_2, \tag{30}$$

where $X_1$ and $X_2$ are homothetic vector fields in the induced geometry on each of the two dimensional non flat submanifolds of constant $q$, $f$ and $t$, $r$, respectively. Now we interested to find homothetic vector fields in the induced geometry on each of the two dimensional non flat submanifolds of constant $q$, $f$ and $t$, $r$. One can easily check that each of the two dimensional non flat submanifolds of constant $t$ and $r$ is of constant curvature. It follows from [7] that homothetic vector fields in the induced geometry on each of the two dimensional non flat submanifolds of constant $t$ and $r$ are

$$\frac{\partial}{\partial f}, \quad \cos f \frac{\partial}{\partial q} - \frac{f'}{f} \sin f \frac{\partial}{\partial f}, \quad \sin f \frac{\partial}{\partial q} + \frac{f'}{f} \cos f \frac{\partial}{\partial f}, \tag{31}$$

which are Killing vector fields. It also follows from [7] that homothetic vector field in the induced geometry on each of the two dimensional non flat submanifolds of constant $q$ and $f$ is $\frac{\partial}{\partial r}$, which is Killing vector field. Hence affine vector fields in the above space-time (29) are Killing vector fields which are given in (6).

**Case A5**

In this case we have $\dot{A} \neq 0$, $\dot{B} \neq 0$, $\dot{A}^2 - 2\ddot{A}A = 0$, $\dot{B}^2 - 2\ddot{B}B = 0$, $\dot{B}^2 + 4B = 0$ and the rank of the $6 \times 6$ Riemann matrix is two. Equations $\dot{A}^2 - 2\ddot{A}A = 0$, $\dot{B}^2 - 2\ddot{B}B = 0$ and $\dot{B}^2 + 4B = 0$ imply that $A(t) = (c_1 t + c_2)^2$ and $B(t) = -t^2 + c_3 t - \frac{1}{4} c_3^2$, where $c_1, c_2, c_3 \in R(c_1 \neq 0)$. Here there exists a unique solution (up to a multiple) $t_a = t_{,a}$ of equation (4) but $t_a$ is not covariantly constant. The line element takes the form

$$ds^2 = -dt^2 + (c_1 t + c_2)^2 dr^2 + (-t^2 + c_3 t - \frac{1}{4} c_3^2)(dq^2 + f^2(q) df^2). \tag{32}$$

Using the above information into the affine equations and after some calculation one find that affine vector fields in this case are Killing vector fields which are given in equation (6).

**Case A6**

Here we have $\dot{A} = 0$, $\dot{B} = 0$ and the rank of the $6 \times 6$ Riemann matrix is one. Equations $\dot{A} = 0$ and $\dot{B} = 0$ imply that $A(t) = c_1$, $B(t) = a$ where $c_1, a \in R \setminus \{0\}$. Here there exist two linearly independent solutions to (4), namely $t_a$ and $r_a$, which are covariantly constant. The line element (after a suitable rescaling of $r$) takes the form

$$ds^2 = -dt^2 + dr^2 + a(d\mathbf{q}^2 + f^2(\mathbf{q})d\mathbf{f}^2). \tag{33}$$

Clearly, the above space-time is seen to be 1+1+2 decomposable and the affine vector fields in this case take the form [2]

$$X = (c_1 t + c_2 r + c_3)\frac{\partial}{\partial t} + (c_4 t + c_5 r + c_6)\frac{\partial}{\partial r} + X', \tag{34}$$

where $c_1, c_2, c_3, c_4, c_5, c_6 \in R$ and $X'$ is a homothetic vector field in each of the two-dimensional submanifolds of constant $t$ and $r$. Now we are interested to find homothetic vector field in the induced geometry on the two-dimensional submanifolds of constant $t$ and $r$, which are given in equation (31). Affine vector fields in this case are

$$X^0 = (c_1 t + c_2 r + c_3), \quad X^1 = (c_4 t + c_5 r + c_6),$$
$$X^2 = c_7 \cos \mathbf{f} + c_8 \sin \mathbf{f}, \quad X^3 = -c_7 \frac{f'}{f}\sin \mathbf{f} + c_8 \frac{f'}{f}\cos \mathbf{f} + c_9, \tag{35}$$

where $c_7, c_8, c_9 \in R$. One can write the above equation (35) after subtracting Killing vector fields as

$$X = (c_1 t + c_2 r + c_3, \, c_4 t + c_5 r + c_6, \, 0, 0). \tag{36}$$

Clearly, the above space-time (33) admits proper affine vector fields.

**Case A7**

In this case the rank of the Riemann matrix is one and we have the conditions $\dot{A} = 0$, $\dot{B} \neq 0$, $\dot{B}^2 - 2\ddot{B}B = 0$ and $\dot{B}^2 + 4B \neq 0$. Equations $\dot{A} = 0$ and

$$\dot{B}^2 - 2\ddot{B}B = 0 \Rightarrow \quad A(t) = c_1, \quad \text{and} \quad B(t) = (c_2 t + c_3)^2, \quad \text{where} \quad c_1, c_2, c_3 \in R($$

$c_1 \neq 0, c_2 \neq 0$). Here there exist two independent solutions $t_a$ and $r_a$ to equation (4). The vector field $r_a$ is covariantly constant whereas $t_a$ is not covariantly constant. The line element after a suitable rescaling of $r$ takes the form

$$ds^2 = -dt^2 + dr^2 + (c_2 t + c_3)^2 (d\mathbf{q}^2 + f^2(\mathbf{q}) d\mathbf{f}^2). \tag{37}$$

Substituting the above information into affine equations, and after lengthy calculation one find that affine vector fields in this case are given in equation (36).

**Case A8**

In this case we have $\dot{A} \neq 0$, $\dot{B} = 0$ $\dot{A}^2 - 2\ddot{A}A = 0$ and the rank of the $6 \times 6$ Riemann matrix is one. Equations $\dot{B} = 0$ and $\dot{A}^2 - 2\ddot{A}A = 0$ imply that $A(t) = (c_1 t + c_2)^2$ and $B(t) = c_3$, where $c_1, c_2, c_3 \in (c_1 \neq 0, c_3 \neq 0)$. Here equation (4) has two linearly independent solutions $t_a$ and $r_a$. The vector fields $t_a$ and $r_a$ are not covariantly constant and the line element is given by

$$ds^2 = -dt^2 + (c_1 t + c_2)^2 dr^2 + c_3 (d\mathbf{q}^2 + f^2(\mathbf{q}) d\mathbf{f}^2). \tag{38}$$

Substituting the above information into affine equations, one finds affine vector fields in this case are given in equation (36).

## SUMMARY

In this paper a study of Kantowski-Sachs and Bianchi type III space-times according to their proper affine symmetry is given. An approach is adopted to study the above space-times by using the rank of the $6 \times 6$ Riemann matrix, holonomy and decomposability and direct integration techniques. From the above study we obtain the following results:

(i) The case when the rank of the $6 \times 6$ Riemann matrix is three and there exists a nowhere zero independent spacelike vector field which is the solution of equation (4) and also covariantly constant. This is the space-time (10) and it admits proper affine vector field (see case A2).

(ii) The case when the rank of the $6\times 6$ Riemann matrix is three or two and there exists a unique nowhere zero independent timelike vector field which is a solution of equation (4) and is not covariantly constant. These are the space-times (24) and (32) and it admits affine vector fields which are Killing vector fields (for details see cases A3 and A5).

(iii) The case when the rank of the $6\times 6$ Riemann matrix is three and there exists a nowhere zero independent vector field which is the solution of equation (4) and is not covariantly constant. This is the space-time (24) and it admits proper affine vector fields (for details see equation (27)).

(iv) The case when the rank of the $6\times 6$ Riemann matrix is two and there exists no solution of equation (4). This is the space-time (29) and it admits affine vector fields which are Killing vector fields (see for details Case A4).

(v) In the case when the rank of the $6\times 6$ Riemann matrix is one there exist two nowhere zero independent vector fields which are solutions of equation (4) and are covariantly constant. This is the space-time (33) and it admits proper affine vector fields (see case A6).

(vi) The case when the rank of the $6\times 6$ Riemann matrix is one and there exist two nowhere zero independent solution of equation (4) but only one independent covariantly constant vector field. This is the space-time (37) and it admits proper affine vector fields (see case A7).

(vii) The case when the rank of the $6\times 6$ Riemann matrix is one and there exist two nowhere zero independent solution of equation (4) but no covariantly constant vector field. This is the space-time (38) and it admits proper affine vector fields (see case A8).

(viii) The case when the rank of the $6\times 6$ Riemann matrix is three and there exists no non-trivial solution of equation (4). This is the space-time (9) and it admits affine vector fields which are Killing vector fields (for details see case A1).

Author's addresses:

Ghulam Shabbir: Faculty of Engineering Sciences, GIK Institute of Engineering Sciences and Technology, Topi, Swabi, NWFP, Pakistan.
Email: shabbir@giki.edu.pk

Nisar Ahmed: Faculty of Electronic Engineering, GIK Institute of Engineering Sciences and Technology, Topi, Swabi, NWFP, Pakistan.